\documentclass[showpacs,amsmath,amssymb,aps,pra,preprint,groupedaddress]{revtex4-1}

\usepackage[latin1]{inputenc}

\begin{document}


\title{Confinement of Fermions via the Spin Connection}


\author{J. B. Formiga}
\email[]{jansen.formiga@uespi.br}
\affiliation{Centro de Ciências da Natureza, Universidade Estadual do Piauí,  64002-150 Teresina, Piauí, Brazil}


\date{\today}

\begin{abstract}
Recently there has been an interest in studying the role that geometry may play in the problem of confining fermions to our four-dimensional spacetime. In general, the focus is on non-Riemannian geometries which possess fields like torsion and non-metricity. In this paper, the degrees of freedom present in the spin connection in a Riemannian manifold are used to confine fermions. It turns out that one can use this connection to trap these particles inside a brane.
\end{abstract}

\pacs{04.50.-h, 04.60.-m, 03.65.Pm }

\maketitle

\section{Introduction}
Due to the fact that we detect only four dimensions in our world, most of the theories and models of physics make use of four-dimensional spacetimes. Nonetheless, 
many physicists have brought their attention to the possibility that our Universe has more than four dimensions. The reason for such an interest in theories with higher dimensions comes from many different perspectives and goals. In string theory, for example, one needs extra dimensions for consistency \cite{Blagojevic:2002du}, while in Kaluza-Klein theory the extra dimension is used to geometrize the electromagnetism \cite{Kaluza:1921tu,Kaluza:1984ws}. Extra dimensions are also used to explain why gravity is weaker than other interactions \cite{ArkaniHamed:1998rs,Antoniadis:1998ig,Randall:1999ee} and to induce matter \cite{Wesson:1999nq}. These example are by themselves sufficient to justify the assumption of extra dimensions. However, all of them need to explain why we do not detect the effects of these extra dimensions. If a model has a compact extra dimension, then one assumes that its radius of compactification is so small that we are not able to detect this dimension yet. In the case of large extra dimensions, one demands that the ordinary matter be confined to our four-dimensional spacetime, which one usually refers to as ``brane''. An interesting approach to this problem, which will be considered here, is to analyze the role that geometry may play to ensure the confinement \cite{Romero:2011qx}.

In Refs. \cite{Romero:2011qx,Dahia:2008zz,Romero:2011zza}, the authors use geometrical fields which appear in either Riemannian or non-Riemannian geometries to confine particles. In \cite{Dahia:2008zz,Romero:2011zza} the authors focus on a classical system, while in Ref. \cite{Romero:2011qx} they approach a quantum one. In the latter case, there are many degrees of freedom which come from the spin connection (sometimes called internal connection) that have not been studied from the perspective of the confinement of fermions. To exemplify the importance that these degrees of freedom may have to this subject, in Sec. \ref{s3}, I present a choice for the spin connection that provides this confinement. Secs. \ref{ss1} and \ref{s2} are devoted to the notation and the Dirac equation. Some final remarks are given in Sec. \ref{fr}.

\section{\label{ss1} Notation}
\subsection{ Vielbein}

One calls ``vielbein''  a set of one-forms $\{\theta^A \}$ such that the metric written in terms of it coincides with the Minkowski metric, i.e., $g=\eta_{AB}\theta^A\otimes\theta^B=(\theta^{(0)})^2-(\theta^{(1)})^2 -\ldots -(\theta^{(n-1)})^2$, where $n$ is the spacetime dimension. Its dual basis is a set of vector fields, which we denote by $\{e_A\}$.  When numbered, the capital Latin indexes will be surrounded by parentheses, like in the metric above.

\subsection{Affine Connection}
Given the vielbein $\theta^A$ and its dual basis $e_A$, the components of the affine connection $\nabla$ in this basis is
\begin{equation}
\omega^A_{\ \ BC} \equiv \theta^A \left(\mathop{\nabla}_{e_B} e_C \right). \label{4042012a}
\end{equation}
The components of $\theta^A$ and $e_A$ in a coordinate basis are denoted by $e^A_{\ \ \mu}$ and $e_A^{\ \ \mu}$, respectively.

\section{Dirac Equation in a Curved Space \label{s2}}
In a five-dimensional Riemannian spacetime, the Dirac equation reads
\begin{equation}
i\gamma^{\mu}\left(\partial_{\mu}+\Gamma_{\mu} \right)\Psi -m\Psi=0, \label{3042012a}
\end{equation}
where
\begin{equation}
\Gamma_{\alpha}=\frac{1}{8}\omega_{A\alpha B} [\gamma^A,\gamma^B]+V_{\alpha} \label{3042012b}
\end{equation}
is the spin connection, and $V_{\alpha}$ belongs to the Clifford algebra $Cl_{3+1}$. 
It is important to emphasize that, for a five-dimensional spacetime, one uses the Clifford algebra $Cl_{3+1}$, since there is no version of it for spacetimes with an odd dimension. Besides, the fifth Dirac matrix is taken to be $\gamma^{(4)}\equiv -\gamma^{(0)}\gamma^{(1)}\gamma^{(2)}\gamma^{(3)}$, where $\gamma^A$ is the Dirac matrix in a certain representation. The value of $\gamma^{(4)}$ in a coordinate basis is $\gamma^{4}(x)=(\varepsilon_{0123}(x))^{-1} \gamma^{(4)} $, where $\varepsilon_{\mu \nu \alpha \beta}=e^A_{\ \ \mu}e^B_{\ \ \nu}e^C_{\ \ \alpha}e^D_{\ \ \beta} \varepsilon_{ABCD}$ is related to the Levi-Civita alternating symbol $\epsilon_{ABCD}$ by $\varepsilon_{ABCD}=\epsilon_{ABCD}$ with $\varepsilon_{(0)(1)(2)(3)}=\epsilon_{0123}=1$.

In terms of the generators of this algebra, the set $\{\mathbb{I},\gamma^A,[\gamma^A,\gamma^B],\gamma^{[A}\gamma^{B}\gamma^{C]},\gamma^{(0)}\gamma^{(1)}\gamma^{(2)}\gamma^{(3)}\}$, we can write   
\begin{eqnarray}
V_{\alpha}=a_{\alpha}\mathbb{I}+a_{\alpha A}\gamma^A+a_{\alpha AB}[\gamma^A,\gamma^B]+
\nonumber \\
a_{\alpha ABC}\gamma^{[A}\gamma^{B}\gamma^{C]}+a\gamma^{(0)}\gamma^{(1)}\gamma^{(2)}\gamma^{(3)}.
\end{eqnarray}

\section{Confinement of Fermions \label{s3}}
The confinement of fermions can be achieved by taking
\begin{equation}
V_{\nu}=\left(f_{,\beta}g_{\nu\alpha}-f_{,\alpha}g_{\nu\beta} \right) \gamma^{[\alpha}\gamma^{\beta]} \label{26012012a}
\end{equation}
and making a suitable choice for the function $f=f(x^4)$. For simplicity, let us assume that the affine connection in the vielbein basis vanishes (Minkowski spacetime with Cartesian coordinates). From this assumption and (\ref{26012012a}), we can write the five-dimensional Dirac equation  in the form
\begin{equation}
i\gamma^{\mu}\partial_{\mu} \Psi + 8if^{\prime}\gamma^4\Psi -m\Psi=0, \label{26012012b}
\end{equation}
where $f^{\prime}=\partial f(x^4)/\partial x^4$. By making the substitution $\Psi=e^{-8f(x^4)}\varphi$, one may rewrite Eq. (\ref{26012012b}) as
\begin{equation}
i\gamma^{\mu}\partial_{\mu} \varphi-m\varphi=0. \label{26012012c}
\end{equation}
Since we are using Cartesian coordinates in Minkowski, we have $\gamma^{\mu}=\gamma^{(\mu)}=constant$. Therefore, it is clear that the solution of Eq. (\ref{26012012c}) can be taken as a wavelike solution of the type
\begin{equation}
\varphi=e^{-ip_{\mu}x^{\mu}}u(p),
\end{equation}
where $p_{\mu}$ is the five-momentum of the particle. It is straightforward to verify that
\begin{equation}
\overline{\Psi}\Psi=e^{-16f(x^4)}\overline{\varphi}\varphi.
\end{equation}
A suitable choice of $f$ can make $\overline{\Psi}\Psi$ go to zero outside a brane characterized by $x^4=x^4_0$. As a result, we manage to confine all the fermions in this brane. As an example, we may take $f(l)=l^2$ ($l\equiv x^4$). In this case, the particles stay inside the brane $l=0$.

\section{Final Remarks \label{fr}} 

It is well-known that the affine connection (\ref{4042012a}) vanishes\footnote{It vanishes only along the curve described by the observer.} when a freely falling observer uses a convenient coordinate system (see, e.g., p. 330 in Ref. \cite{Gravitation}). In addition, the spin connection (\ref{3042012b}) must also vanish in the absence of non-gravitational interactions so that the equivalence principle holds.  Therefore, the matrix $V_{\mu}$ in (\ref{3042012b}) has to be related to a non-gravitational interaction  to ensure that this principle holds, unless $V_{\mu}$ vanishes under the same conditions as the affine connection does. Since $V_{\mu}$ has a geometrical character, relating it to an interaction means geometrizing this interaction (as an example of such geometrization, see Ref. \cite{Novello:1973jd}). Nonetheless, we only need to guarantee the validity of the equivalence principle in four dimensions. If we choose $f(x^4)$ in such a way that the particles are trapped to the brane $x^4=0$ and $\left(df(l)/dl \right)|_{l=0}=0$, then $V_{\mu}$ would vanish in our world (see the example in the previous section).

\end{document}